\documentclass[letter]{IEEEtran}

\usepackage{pgfplots}
\usepackage{hyperref}
\usetikzlibrary{shapes.multipart,intersections}
\usepackage{cite}
\usepackage{amsmath,amssymb,amsfonts,steinmetz}
\usepackage{algorithmic}
\usepackage{graphicx}
\usepackage{textcomp}
\usepackage{xcolor}
\def\BibTeX{{\rm B\kern-.05em{\sc i\kern-.025em b}\kern-.08em
    T\kern-.1667em\lower.7ex\hbox{E}\kern-.125emX}}

\usepackage{tikz}
\usetikzlibrary{calc}
\makeatletter
\newcommand{\gettikzxy}[3]{%
  \tikz@scan@one@point\pgfutil@firstofone#1\relax
  \edef#2{\the\pgf@x}%
  \edef#3{\the\pgf@y}%
}
\makeatother
\usepackage[draft]{todonotes}   % notes showed

\usepackage[T1]{fontenc}
\usepackage[latin9]{inputenc}
\usepackage{bm}
\usepackage{amsmath}
\usepackage{amssymb}
\usepackage{stmaryrd}
\usepackage{babel}
\usepackage{graphics, graphicx}
\usepackage{xcolor}
\usepackage{gensymb}
\usepackage{cite}
\usepackage{enumitem}
\usepackage{url}
\newtheorem{rem}{Remark}

\begin{document}
\title{Localization via Multiple Reconfigurable Intelligent Surfaces Equipped with Single Receive RF Chains}
\author{George C. Alexandropoulos,~\IEEEmembership{Senior~Member,~IEEE}, Ioanna Vinieratou,~\IEEEmembership{Student~Member,~IEEE}, \\and Henk Wymeersch,~\IEEEmembership{Senior~Member,~IEEE}
\thanks{This work has been supported by the EU H2020 RISE-6G project under grant number 101017011.}
\thanks{G. C. Alexandropoulos and I. Vinieratou are with the Department of Informatics and Telecommunications,
National and Kapodistrian University of Athens, Greece (e-mails: \{alexandg, en2180001\}@di.uoa.gr).}
\thanks{H. Wymeersch is with the Department of Electrical Engineering, Chalmers University of Technology, Sweden (e-mail: henkw@chalmers.se).}
\vspace{-0.3cm}
}

\maketitle
\begin{abstract}
The extra degrees of freedom resulting from the consideration of Reconfigurable Intelligent Surfaces (RISs) for smart signal propagation can be exploited for high accuracy localization and tracking. In this paper, capitalizing on a recent RIS hardware architecture incorporating a single receive Radio Frequency (RF) chain for measurement collection, we present a user localization method with multiple RISs. The proposed method includes an initial step for direction estimation at each RIS, followed by maximum likelihood position estimation, which is initialized with a least squares line intersection technique. Our numerical results showcase the accuracy of the proposed localization, verifying our theoretical estimation analysis.  
\end{abstract}

\vspace{-4mm}
\section{Introduction}
Reconfigurable Intelligent Surfaces (RISs) \cite{huang2019reconfigurable} are lately gaining increased interest as a cost- and power-efficient means to enable programmable wireless signal propagation environments \cite{rise6g}. Due to their minimal hardware footprint, they are envisioned to coat available surfaces or objects in the wireless medium, offering extra degrees of freedom for diverse communication, localization, and sensing improvements. RISs are artificially planar structures usually consisting of multiple unit elements (of half- or even sub-wavelength inter-element spacing), whose reflection behavior can be adjusted to finite discrete states \cite{huang2019holographic}. The dynamic configuration of RISs is handled by dedicated controllers connected to them, while lately RIS hardware architectures including small numbers of Radio Frequency (RF) chains are being proposed \cite{George_RIS_2020,shlezinger2020dynamic,HRIS}, facilitating several RIS-enabled network management tasks.

Among the possible functionalities, where RISs can have a significant contribution, belong the high-accuracy localization. This feature constitutes one of the key requirements for fifth Generation (5G), and beyond, wireless networks \cite{Positioning_5G_NR}. An overview of the main challenges and opportunities for localization and mapping with RIS-empowered wireless systems is provided in \cite{wymeersch2019radio}. Considering passive RISs with no active RF chains and quantized phase profiles in \cite{he2020large}, the Cram\'{e}r-Rao lower bound for RIS-based positioning in millimeter Wave (mmWave) Multiple-Input Multiple-Output (MIMO) systems was presented.
% In \cite{hu2018beyond}, authors have derived the Fisher-information and Cramer-Rao Lower Bounds (CRLBs), for positioning with continuous large intelligent surfaces (LIS). 
For the same systems, an adaptive hierarchical codebook for positioning and data transmission was designed in \cite{he2020adaptive}. A supervised learning approach for wave fingerprinting was proposed in \cite{alexandg_2021} for localizing non-cooperative objects in rich scattering RIS-empowered wireless systems. Simultaneous localization and mapping enabled by passive RISs was proposed in \cite{yang2021wireless}. In \cite{keykhosravi2020siso}, a joint three-dimensional localization and synchronization approach for a single-input single-output system empowered by an RIS was designed. In addition to the above works on RISs without RF chains, \cite{shaban2020nearfield} exploited 
wavefront curvature for positioning, by utilizing RIS-based lenses with a single RF chain for reception.

In this paper, motivated by the recent interest in RISs with basic sensing capability \cite{George_RIS_2020,HRIS}, we design a novel user localization method realized with multiple RISs, each equipped with a single Receive (RX) RF chain. Assuming that each meta-atom element of an RIS is coupled with a waveguide and that all waveguide outputs are fed to the RF chain \cite{George_RIS_2020,shlezinger2020dynamic}, we present the following contributions: \emph{i)} We design an Angle-of-Arrival (AoA) estimation technique for single-RX-RF RISs, which is based on collected measurements using multiple RIS phase profiles. \emph{ii)} We capitalize on the AoA estimations from multiple RISs for a source of interest and present a Maximum Likelihood (ML) localization algorithm that is based on a line intersection technique. \emph{iii)} We also derive the Position Error Bounds (PEBs) for the unknown parameters and provide numerical results that demonstrate the accuracy of the proposed localization for various system parameters, including different RIS codebooks, different quantization of RIS phase profiles, and different RIS placements, as well as under near- and far-field conditions. 

\subsubsection*{Notations} Vectors and matrices are denoted by boldface lowercase and boldface capital letters, respectively. The transpose, Hermitian transpose, and inverse of $\mathbf{A}$ are denoted by $\mathbf{A}^{\rm T}$, $\mathbf{A}^{\rm H}$, and $\mathbf{A}^{-1}$, respectively, while $\mathbf{I}_{n}$ ($n\geq2$) is the $n\times n$ identity matrix and $\mathbf{0}_{n\times m}$ ($n,m\geq1$) is an $n\times m$ matrix with zeros. ${\rm Tr}\{\mathbf{A}\}$ represents $\mathbf{A}$'s trace and $[\mathbf{A}]_{i,j}$ denotes its $(i,j)$-th element, while $[\mathbf{A}]_{:,i:j}$ is a submatrix of $\mathbf{A}$ including all rows of $\mathbf{A}$ and columns from the $i$-th up to the $j$-th. The Euclidean norm of $\mathbf{a}$ is denoted by $\left\|\mathbf{a}\right\|$ and $\odot$ is the matrix Hadamard product. $\mathbb{R}$ and $\mathbb{C}$ represent the real and complex number sets, respectively, and $\Re\left\{\cdot\right\}$ gives the real part of a complex matrix. $\mathbb{E}\{\cdot\}$ is the expectation operator, and $\mathbf{x}\sim\mathcal{CN}(\mathbf{a},\mathbf{A})$ indicates a complex Gaussian random vector with mean $\mathbf{a}$ and covariance matrix $\mathbf{A}$. Finally, $x\sim\mathcal{U}(a,b)$ denotes a uniformly distributed random variable in the interval $[a,b]$, and $\jmath\triangleq\sqrt{-1}$ is the imaginary unit.
\vspace{-2mm}
\section{System Model}\label{sec:System_Model}

%\subsection{System Model}
\begin{figure}[!t]
	\centering
	\includegraphics[trim=7cm 1.5cm 1.8cm 5cm, width=1\columnwidth]{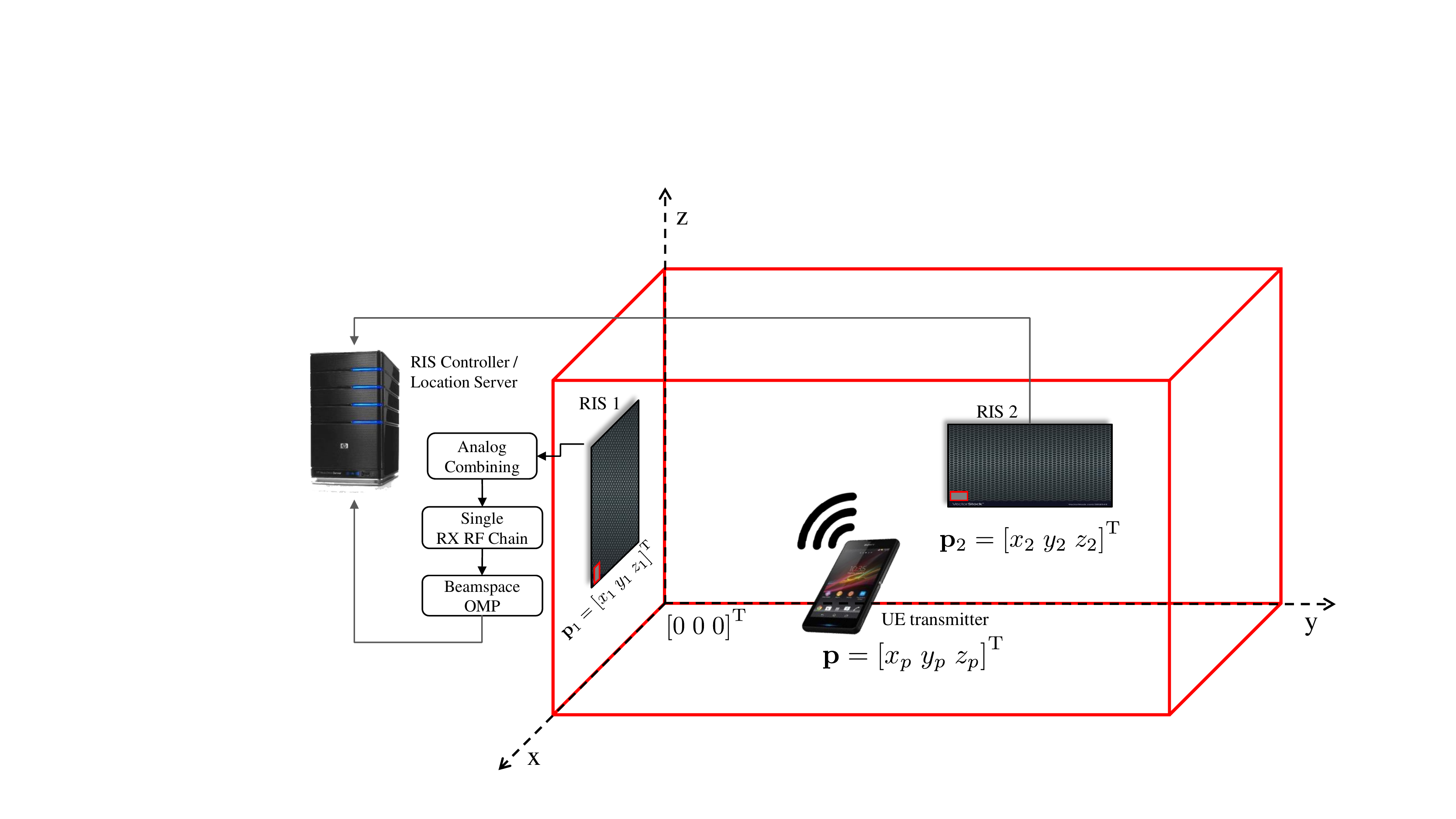}
	  \caption{The considered indoor setup with $M$ single-RX-RF RISs, each with $L$ elements, for the localization of a transmitting user at the unknown position $[x_p\,y_p\,z_p]^{\rm T}$. The position $[x_m\,y_m\,z_m]^{\rm T}$, with $m=1,2,\ldots,M$, of the reference unit element at each $m$-th RIS is assumed to be known.}\vspace{-0.4cm}
		\label{fig:system_setup}
\end{figure}
We consider the indoor environment of Fig.~\ref{fig:system_setup} comprising $M\ge2$  RISs, which are attached to the room's walls and are themselves planar and perpendicular to the floor. Each RIS consists of $L$ phase-tunable meta-atom elements and is implemented with the single-RX-RF architecture of \cite{George_RIS_2020}. There is a global coordinate system, where the reference point of each $m$-th RIS, with $m=1,2,\ldots,M$, is given by $\mathbf{p}_m\triangleq\left[x_m\;y_m\;z_m\right]^{\rm T}$ and its orientation (in terms of the azimuth angle) is represented by the parameter $\beta_m$. All RISs are assumed to be connected to the same central controller. Finally, a transmitting single-antenna user (the source) is located at the unknown position $\mathbf{p}\triangleq\left[x_p\;y_p\;z_p\right]^{\rm T}$. {It is noted that a similar system model can be considered for outdoor environments, where RISs can be deployed for coating, for example, building facades. This paper's localization method can be equally applied for such a system model.}

%\subsection{Signal Model}
{We assume that the user broadcasts a pilot symbol $s$ with constant transmit power $P$. This symbol is received $T$ times by each RIS, where during each repetition a different RIS phase profile is used. Under far-field signal propagation and in the presence of $C_{m}$ distinct channel paths, the observation during the $t$-th reception slot} ($t=1,2,\ldots,T$) at each $m$-th RIS's RX RF chain output can be mathematically expressed as follows:
{\begin{align}
    y_{m,t}\triangleq \mathbf{u}_{m,t}^{\rm H}\sum_{c=1}^{C_{m}} h_{m,c}\boldsymbol{\alpha}\left(\phi_{m,c},\theta_{m,c}\right)s+\mathbf{u}_{m,t}^{\rm H}\mathbf{w}_{m,t},  \label{eq:ObservedSignalmthRIS}
\end{align}
where $h_{m,c}\triangleq\sqrt{P_{L_{m,c}}}\exp\left(\jmath\varphi_{m}\right)$ $\forall$$c=1,2,\ldots,C_m$ includes the gain of the $c$-th signal propagation path with parameter $P_{L_{m,c}}\triangleq\lambda^2/\left(4\pi r_{m,c}\right)^2$ denoting the free-space pathloss, where $\lambda$ is the signal wavelength. Without loss of generality, we assume that the $c=1$ channel path represents the Line-Of-Sight (LOS), hence, its pathloss $P_{L_{m,1}}$ depends on the Euclidean distance $r_{m,1}\triangleq\|\mathbf{p}_m-\mathbf{p}\|$; each distance $r_{m,c}$ for $c\geq2$ is defined similarly considering the position of the corresponding scatterer. In the expression for $h_{m,c}$, $\varphi_m\sim\mathcal{U}(0,2\pi)$ denotes a global phase offset accounting for the lack of phase synchronization between the user and the RF chain of the $m$-th RIS.} The vector $\mathbf{u}_{m,t}\in\mathbb{C}^{L\times 1}$ is the $t$-th phase configuration (among the total $T$ used) of the $m$-th RIS. The vector $\mathbf{w}_{m,t}\in\mathbb{C}^{L\times 1}$ in \eqref{eq:ObservedSignalmthRIS} represents the Additive White Gaussian Noise (AWGN) that is distributed as $\mathcal{CN}(\mathbf{0},\rho\mathbf{I}_L)$. {Finally, the spatial response vector $\boldsymbol{\alpha}\left(\phi_{m,c}, \theta_{m,c}\right)\in\mathbb{C}^{L\times1}$ for the azimuth and elevation AoAs $\phi_{m,c}\in\left[0,2\pi\right]$ and $\theta_{m,c}\in\left[0,\pi\right]$, respectively, of the user transmitted signal via multipath propagation, with respect to the coordinate system having as origin the point $\mathbf{p}_m$ (i.e., the reference point of the $m$-th RIS), is given for $\ell=1,2,\ldots,L$ as follows:
\begin{equation}\label{eq:Spatial_response}
    \left[\boldsymbol{\alpha}\left(\phi_{m,c},\theta_{m,c}\right)\right]_\ell \triangleq  \exp\left(-\jmath\mathbf{q}^{\rm T}_{m,\ell}\mathbf{k}(\phi_{m,c},\theta_{m,c})\right),
\end{equation}
where $\mathbf{q}_{m,\ell}\in\mathbb{R}^{3\times1}$ denotes the position of the $\ell$-th element
%\footnote{It is noted that the coordinates of each $\ell$-th element of each $m$-th RIS can be easily obtained from the known reference point $\mathbf{p}_m$. For example, consider the $m$-th RIS in Fig.~\ref{fig:system_setup} that lies on the $yz$ plane and has a uniform placement of its $L$ elements in a square grid of $N$ elements per side (i.e., $L=N\times N$) with inter-element distance $d$. We also assume that those elements are ordered as $1,2,\ldots,L$, starting from the reference point $\mathbf{q}_{m,1}$ with coordinates $(x_m,y_m,z_m)$ and the rest $N-1$ elements in the same first line, then the second line with the next set of $N$ elements having the point $\mathbf{q}_{m,N+1}$ with coordinates $(x_m,y_m+(N-1)d,z_m)$ as the line's first element, and so on. Using the latter ordering and the coordinate system of Fig.~\ref{fig:system_setup}, the coordinates of the $\ell$-th element of the $m$-th RIS, i.e., of $\mathbf{q}_{m,\ell}$, needed in \eqref{eq:Spatial_response} are $(x_m,y_m+\gamma d,z_m+\delta d)$, where $\delta$ and $\gamma$ are respectively the quotient and remainder of the division of $\ell-1$ with $N$, i.e., $\ell-1=N\delta+\gamma$.} 
of the $m$-th RIS and $\mathbf{k}(\phi_{m,c}\theta_{m,c})\in\mathbb{R}^{3\times1}$ is the wavevector at these respective AoAs, which is mathematically defined as:
\begin{align}
\mathbf{k}(\phi_{m,c},\theta_{m,c}) \triangleq -\frac{2 \pi}{\lambda}\left[\begin{array}{c}
\sin\theta_{m,c}\cos\phi_{m,c}\\
\sin\theta_{m,c}\sin\phi_{m,c}\\
\cos\theta_{m,c}
\end{array}\right]. 
\end{align}}

%\subsection{RIS Phase Profiles}

The phase profiles of each $m$-th RIS are selected from a set $\mathcal{U}$ with cardinality $K$ such that $\mathbf{u}_{m,k}\in \mathcal{U}$ $\forall$$k=1,2,\ldots,K$. Note that, in general, $K$ may be larger or smaller than $T$.  
Assuming that $b$ is the phase resolution in bits per RIS phase-tunable unit element, we consider the $2^{b}$-element discrete set $\mathcal{F}\triangleq\{\exp\left(\jmath2^{1-b}\pi f\right)\}^{2^{b}-1}_{f=0}$ for the elements $[\mathbf{u}_{m,k}]_\ell$ $\forall m,k,\ell$, 
which results in a total of $K=2^{bL}$ phase profiles. %We constrain each of these profiles as $\Vert \mathbf{u}_{m,k}\Vert^2=1$. 
%\Henk{Ioanna: do the RIS phase profiles satisfy $\Vert \mathbf{u}\Vert^2=1$ or $\Vert \mathbf{u}\Vert^2=L$?} \George{They are then normalized as in \eqref{eq:dft} to have $\Vert \mathbf{u}\Vert^2=1$.}

\vspace{-2mm}
\section{Proposed RIS-Enabled Localization Method}
The proposed localization method comprises two stages: \textit{i}) AoA estimation {of the LOS channel component} in azimuth and elevation from each RIS; and \textit{ii}) user position estimation, by fusing the AoA measurements {of the LOS channel} from the individual RISs. {For the former stage,
%various conventional high-resolution algorithms can be used (e.g., beamspace MUltiple SIgnal Classification (MUSIC) \cite{chuang15high}), or more efficiently, compressed-sensing-based approaches (e.g., based on the Orthogonal Matching Pursuit (OMP) \cite{wu2013Support,wang2015Support}). In this paper, we present an OMP-based beamspace technique 
we apply a beamspace version of the Orthogonal Matching Pursuit (OMP) algorithm \cite{wu2013Support,wang2015Support}, which will be shown to perform accurate AoA estimations with relatively small values for $T$.
%for the direction finding problem at each single-RX-RF RIS in multipath channel conditions, which can perform accurate AoA estimations with even only $S=1$ observation per transmission slot. 
The second stage for user localization is based on an ML approach, which is initialized with a line intersection technique that is based on the Least Squares (LS) criterion.} In terms of system architecture, we assume that the baseband measurements at the outputs of each RIS's RX RF chain are collected by a central controller or location server. This device can then estimate the AoA of the transmitted signal's {LOS component} at each RIS, and then, fuse all those $M$ AoA estimates to obtain the user position estimation, as will be detailed in the next section. It is noted that the AoA estimation can be performed at any RIS side at the cost of basic storage and computing capability.

\vspace{-2mm}
\subsection{AoA Estimation from Each Single-RX-RF RIS}
{By stacking the $T$ phase profiles for a symbol observation in the $L\times T$ matrix $\mathbf{U}_m\triangleq[\mathbf{u}_{m,1}\,\mathbf{u}_{m,2}\,\cdots\,\mathbf{u}_{m,T}]$, we obtain the following $T$-element column vector using expression \eqref{eq:ObservedSignalmthRIS}: 
\begin{align}\label{eq:Rx_signal2}
\mathbf{y}_{m} \triangleq \mathbf{U}_m^{\rm H}\sum_{c=1}^{C_m}h_{m,c}\boldsymbol{\alpha}\left(\phi_{m,c},\theta_{m,c}\right)s+\tilde{\mathbf{w}}_{m},
	\end{align}
where $\tilde{\mathbf{w}}_{m}\triangleq[\mathbf{u}_{m,1}^{\rm H}\mathbf{w}_{m,1}\,\mathbf{u}_{m,2}^{\rm H}\mathbf{w}_{m,2}\,\ldots\,\mathbf{u}_{m,T}^{\rm H}\mathbf{w}_{m,T}]^{\rm{T}}$.} {We next use the matrix $\boldsymbol{\Psi}\triangleq\mathbf{U}_m^{\rm H}\mathbf{A}\in\mathbb{C}^{T\times J}$ with $J\gg T$, where $\mathbf{A}\triangleq[\mathbf{a}(\bar{\phi}_1,\bar{\theta}_1)\,\mathbf{a}(\bar{\phi}_2,\bar{\theta}_2)\,\cdots\,\mathbf{a}(\bar{\phi}_{J},\bar{\theta}_{J})]\in\mathbb{C}^{L\times J}$ is a dictionary matrix with the spatial response vectors at the azimuth and elevation AoA pairs $(\bar{\phi}_j,\bar{\theta}_j)$ $\forall$$j=1,2,\ldots,J$, to approximate the received symbol vector at each $m$-th RIS as:
\begin{equation}\label{eq:CS}
 \mathbf{y}_{m}\approx\boldsymbol{\Psi}\mathbf{x}_{m}+\tilde{\mathbf{w}}_{m}, 
\end{equation}
where $\mathbf{x}_{m}\in\mathbb{C}^{J\times 1}$ is an approximately $C_{m}$-sparse vector  including the gains of all channel paths via which the user's signal reaches the RIS. The latter formulation enables the deployment of compressed sensing tools for estimating $\mathbf{x}_{m}$, i.e., the AoAs and gains of all channel paths. In this paper, we deploy the OMP algorithm \cite{wu2013Support} to estimate $\phi_{m,1}$ and $\theta_{m,1}$ of the LOS channel, which will be then used for position estimation. This estimation can be obtained by running OMP with sparsity level $1$, when LOS is the strongest channel component. Otherwise, the support of $\mathbf{x}_{m}$ needs to be estimated, and then combined with a data association approach for estimating the LOS component.}

\begin{rem}
To recover the LOS component with the presented OMP, the measurement matrix $\boldsymbol{\Psi}$ needs to satisfy the first-order restricted isometry property \cite{wang2015Support}. One option for meeting this property is when $\boldsymbol{\Psi}$ is the Discrete Fourier Transform (DFT) matrix. In this paper, given the spatial response dictionary $\mathbf{A}$, we choose $\mathbf{U}_m$ as the DFT matrix, i.e., it must hold $T=L$, as well as a partial DFT matrix consisting of $T<L$ columns from the DFT matrix {\cite{candescompressed2011}}. For both cases, the elements of $\mathbf{U}_m$ for $\ell=1,2,\dots,L$ and $t=1,2,\dots,T$ are given by the following expression:
\begin{equation}\label{eq:dft}
 [\mathbf{U}_m]_{\ell,t}=\frac{1}{\sqrt{L}}\exp\Big(-\frac{\jmath2\pi}{L}(\ell-1)(t-1)\Big).
\end{equation}
\end{rem}

{
\begin{rem}
The AoA estimation performance of OMP can be improved by increasing the number of transmissions $T$, which proportionally increases the integrated SNR.% of signal averaging. At the cost of larger observation periods, $S$ vectors $\mathbf{y}_{m,i}$ $\forall$$i=1,2,\ldots,S$ can be collected to formulate \eqref{eq:CS}'s new observation vector~$\bar{\mathbf{y}}_{m}\triangleq S^{-1}\sum_{i=1}^S\mathbf{y}_{m,i}$.
\end{rem}}

%The goal of OMP is to estimate $\mathbf{\bar{x}}_{m}$ by iteratively finding the column of $\Psi$ having the maximum correlation with the current residual $r$  and then updating $r$ by substracting its contribution   $r=y-\Psi\mathbf{\bar{x}}_{m}$. The stopping rule for OMP can be either the number of iterations equal to the sparsity of $\mathbf{x}_{m}$, or a threshold for $r$ defined by a-priori knowledge on $\tilde{\mathbf{w}}_{m,i}$.}

\vspace{-2mm}
\subsection{Centralized ML-Based Position Estimation}
Given the known coordinates $\mathbf{p}_m=\left[x_m\;y_m\;z_m\right]^{\rm T}$ of the reference point at the $m$-th RIS as well as {the estimates $\hat{\phi}_{m,1}$ and $\hat{\theta}_{m,1}$ of the LOS AoAs $\phi_{m,1}$ and $\theta_{m,1}$}, the line where the single source lies can be estimated for each $m$-th RIS. Each point $\boldsymbol{\xi}\triangleq\left[x\;y\;z\right]^{\rm T}$ in this line can be obtained for $k\in\mathbb{R}$ as: 
	\begin{equation}
\boldsymbol{\xi}=\mathbf{p}_m+k\mathbf{v}_m,
	\end{equation}
where $\mathbf{v}_m\in\mathbb{R}^{3\times1}$ is the direction vector, which is defined as 
{\begin{equation}
\mathbf{v}_m\triangleq\left[\begin{array}{c}
                       \sin{(\hat\theta_{m,1})}\cos{(\hat\phi_{m,1}+\beta_{m,1})}\\
                       \sin{(\hat\theta_{m,1})}\sin{(\hat\phi_{m,1}+\beta_{m,1})}\\
                       \cos{(\hat\theta_{m,1})}
                        \end{array}\right].
                        \end{equation}}
By assuming that the controller in the considered system model possesses all $M$ $\mathbf{p}_m$'s and calculates all $M$ angle pairs {$(\hat{\phi}_{m,1},\hat{\theta}_{m,1})$}, it is proposed to compute the estimation for the unknown position $\mathbf{p}$, as follows. The sum of the squared distances between the source and each of the $M$ estimated lines is given as a function of $\mathbf{p}$ by the expression \cite{dupre1992distance}:
	\begin{equation}\label{eq:LS_Distances}
D(\mathbf{p})\triangleq\sum_{m=1}^{M}\left(\mathbf{p}_m-\mathbf{p}\right)^{\rm T}\mathbf{B}_m\left(\mathbf{p}_m-\mathbf{p}\right),
	\end{equation}
where $\mathbf{B}_m\triangleq\mathbf{I}_3-\mathbf{v}_m\mathbf{v}_m^{\rm T}$. The estimation of the unknown $\mathbf{p}$ that minimizes this LS problem can be easily obtained as 
 \begin{equation}\label{eq:intersection}
\hat{\mathbf{p}}_{\text{LS}} =  \Big(\sum_{m=1}^{M}\mathbf{B}_m\Big)^{-1}\Big(\sum_{m=1}^{M}\mathbf{B}_m\mathbf{p}_m\Big).
 \end{equation}

The Root Mean Squared Error (RMSE) using the LS-based intersection of skew lines can be, in general, larger than the PEB, since all AoA estimates are treated equally. This fact may result in poor position estimation performance, if any or some of the $M$ AoA estimates is of poor quality. We next consider uncertainty in the AoA estimates to refine the position estimate. In particular, the estimation of AoAs for each $m$-th RIS can be expressed in the following form:
{\begin{equation}
\hat{\mathbf{{y}}}_m\left(\mathbf{p}\right)\triangleq\begin{bmatrix}
\hat{\theta}_{m,1}\\
\hat{\phi}_{m,1}
\end{bmatrix} = \begin{bmatrix}
\theta_{m,1}\\
\phi_{m,1}
\end{bmatrix}+\mathbf{z}_m=f_m\left(\mathbf{p}\right) + \mathbf{z}_m,
 \end{equation}
where $f_m\left(\mathbf{p}\right)$ is a $2\times 1$ vector containing the azimuth and elevation angles with respect to the $m$-th RIS; this is a non-linear function of $\mathbf{p}$.} In addition, $\mathbf{z}_m\sim\mathcal{N}(0,\mathbf{G}_m)$ is a $2\times 1$ vector modeling the estimation errors for the AoAs. Given this uncertainty model, the ML estimation of the single transmitting user position, $\hat{\mathbf{p}}_{\rm ML}$, is obtained %for $\mathbf{G}_m\triangleq\mathbf{J}\left(\phi_m,\theta_m\right)$ 
as follows:
\begin{equation}\label{eq:ML}
\begin{split}
    \hat{\mathbf{{p}}}_{\rm ML}\!\!\triangleq\!\operatorname*{argmin}_{\mathbf{p}}\!  \sum^M_{m=1}[\hat{\mathbf{{y}}}_m\left(\mathbf{p}\right)-f_m\left(\mathbf{p}\right)]^{\rm{T}}\mathbf{G}^{-1}_m[\hat{\mathbf{y}}_m\left(\mathbf{p}\right)-f_m\left(\mathbf{p}\right)].
   \end{split}
\end{equation}
This optimization problem does not possess a closed-form solution. We hence propose to solve it via any gradient-descent method using the LS-based position estimation in \eqref{eq:intersection}, as the initialization step. To determine each of the $2\times2$ covariance matrices $\mathbf{G}_m$ in \eqref{eq:ML} and establish a fundamental performance bound, we will next rely on a Fisher information analysis. 

\vspace{-4mm}
\subsection{Fisher Information Analysis}
\subsubsection{LOS Channel Parameters}
{Let the vector $\bm{\eta}_m\triangleq\left[\sqrt{P_{L_{m,1}}}\;\varphi_m\;\phi_{m,1}\;\theta_{m,1}\right]^{\rm T}\in\mathbb{R}^{4\times1}$ include the unknown system parameters in \eqref{eq:ObservedSignalmthRIS}, referring to the received signal model via the LOS component at each $m$-th RIS. The $4\times4$ Fisher Information Matrix (FIM) for this unknown vector, based on the observation model in expression \eqref{eq:ObservedSignalmthRIS} when excluding the non-LOS channel paths,
%\footnote{{The signal model considered for the FIM derivation ignores the secondary channel paths, resulting in an overestimation of the PEB performance.}}, 
is defined as follows \cite{Kay97}:}
\begin{equation}\label{eq:FIM_Initial}
\mathbf{J}\left(\bm{\eta}_m\right)=\frac{2}{\rho}\sum_{t=1}^T\Re\left\{\left(\frac{\partial\mu_{m,t}}{\partial\bm{\eta}_m}\right)^{\rm H}\frac{\partial\mu_{m,t}}{\partial\bm{\eta}_m}\right\},
\end{equation}
where {the function $\mu_{m,t}$ of the unknown vector $\bm{\eta}_m$ is defined as $\mu_{m,t} \triangleq \sqrt{P_{L_{m,1}}}\exp\left(j\varphi_{m}\right)\mathbf{u}_{m,t}^{\rm H}\boldsymbol{\alpha}\left(\phi_{m,1},\theta_{m,1}\right)s$, representing} the noiseless received signal at the $t$-th slot at the $m$-th RIS. Using Schur's complement, the Fisher information for the unknown AoAs {$\phi_{m,1}$ and $\theta_{m,1}$} can be computed as:
{\begin{align}
%\begin{split}
& \mathbf{G}^{-1}_m = \mathbf{J}\left(\phi_{m,1},\theta_{m,1}\right)=\left[\mathbf{J}(\bm{\eta}_{m})\right]_{3:4,3:4} \notag \\
& -\left[\mathbf{J}(\bm{\eta}_{m})\right]_{3:4,1:2}\left[\mathbf{J}(\bm{\eta}_{m})\right]_{1:2,1:2}^{-1}\left[\mathbf{J}(\bm{\eta}_{m})\right]_{1:2,3:4}.
\label{eq:fim_schur}
%\end{split}
\end{align}}
%which can be used as a replacement for $\mathbf{G}^{-1}_m$ in \eqref{eq:ML}. 

\subsubsection{Positioning}
The Fisher information for each $m$-th RISs can be used for expressing the $3\times3$ FIM for the unknown position $\mathbf{p}$ as follows \cite{Kay97}:
{\begin{equation}\label{eq:fim_position}
\mathbf{J}\left(\mathbf{p}\right)=\sum_{m=1}^M\mathbf{T}_m\mathbf{J}\left(\phi_{m,1},\theta_{m,1}\right)\mathbf{T}_m^{\rm T},
\end{equation}
where $\mathbf{T}_m\in\mathbb{R}^{3\times2}$ denotes the Jacobian $\mathbf{T}_m\triangleq[\frac{\partial\theta_{m,1}}{\partial\mathbf{p}} \frac{\partial\phi_{m,1}}{\partial\mathbf{p}}]$}. The derivatives needed for this Fisher information of the unknown AoAs and source position are given in the Appendix. 

The FIM for the unknown source position $\mathbf{p}$ can be finally used for computing the PEB, as follows:
\begin{equation}\label{eq:peb}
{\rm PEB}\triangleq\sqrt{{\rm Tr}\left\{\mathbf{J}^{-1}\left(\mathbf{p}\right)\right\}}\leq\sqrt{\mathbb{E}\left\{\left\|\hat{\mathbf{ p}}-\mathbf{p}\right\|^2\right\}},
\end{equation}
which can serve as a lower bound for the RMSE of the position estimation $\mathbf{\hat p}$ for $\mathbf{p}$, when considering the model of Section~\ref{sec:System_Model}.
%After (16) you could mention that this is still a valid bound in the presence of multipath, since the multipath will only degrade RSME performance, thus the bound is still a lower bound
\begin{rem}
The bound in \eqref{eq:peb} is valid even in the presence of multipath, since in this scenario, multipath cannot improve positioning RMSE without data association among the RISs. 
\end{rem}

\vspace{-2mm}
\section{Numerical Results}
%In this section, we investigate the performance of the proposed positioning method over LOS wireless channels, varying the placement of the RISs and their total number $M$, as well as the total number $L$ of their elements and the form of the deployed phase profiles together with their number $T$ used for measurements' collection from each RIS.
\vspace{-1mm}
\subsection{Simulated Scenario}
 We have considered a cubic room with dimensions $10\,{\rm m}\times 10\,{\rm m}\times 10\,{\rm m}$ and single-RX-RF RISs of half-wavelength, i.e., $\lambda/2$, inter-element spacing operating at $30$ GHz with the noise floor $\rho$ in \eqref{eq:ObservedSignalmthRIS} set to $-79$ dBm. Unless otherwise stated, we have assumed transmit power of $P=10$ dBm, $M=3$ RISs of size $L=8\times 8$, and $T=L$ RIS phase profiles resulting from the DFT beams. The RISs were considered placed on the room's vertical walls as depicted in Fig.~\ref{fig:system_setup}. We have generated a multipath channel with the same number of paths for all RISs (specifically, $C_{m}=3$ $\forall$$m$), comprising a LOS and two non-LOS components with $20$ dB power ratio, using both the far-field model in \eqref{eq:Spatial_response} for the steering vectors as well as a near-field model. For the latter model, we have used {the following formula for the LOS component (the non-LOS channel components were modeled similarly):
\begin{equation}\label{eq:Spatial_response2}
    \left[\boldsymbol{\alpha}\left(\phi_{m,1},\theta_{m,1}\right)\right]_\ell \triangleq  \exp\left(-\jmath2 \pi r_{m,1,\ell}/\lambda\right),
\end{equation}
where $r_{m,1,\ell}\triangleq\Vert \mathbf{p}- \mathbf{p}_{m,\ell}\Vert $} with $\mathbf{p}_{m,\ell}\in\mathbb{R}^{3\times1}$ being the absolute coordinate of the $\ell$-th element of the $m$-th RIS, which can be easily obtained from $\mathbf{q}_{m,\ell}$. It is noted that the proposed positioning techniques in Section~III were derived using \eqref{eq:Spatial_response}, and here, their performance over the multipath channel model resulting from \eqref{eq:Spatial_response2} is also investigated.

\begin{figure}

	\centering
	\input{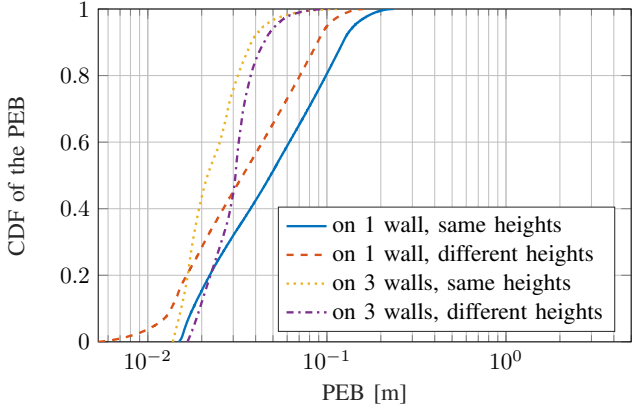}

    \caption{The empirical CDF of the PEB for various source positions with $z_p=5$ ${\rm m}$ and $P=10$ dBm,  as well as $M=3$ RISs each with $L=8\times8$ unit elements. Different RIS placements on the room's walls were considered.} \vspace{-4mm}
    	\label{fig:cdf_plot}
\end{figure}

%\begin{figure}[!t]
%	\centering
%	\includegraphics[width=0.53\textwidth]{Figures/peb_cdf_plot.pdf}
%	  \caption{The empirical CDF of the PEB for various source positions with $z_p=5$ ${\rm m}$ and $P=10$ dBm, as well as $M=3$ RISs each having $L=8\times8$ elements. Different placements of the RISs in one or three of the room's walls were considered.}\vspace{-0.2cm}
%	  		\label{fig:cdf_plot}
%\end{figure}
%, where $\rho$ denotes the noise floor in dBm expressed as $\rho=-174 + {\rm NF}+ \rm{10log10}(BW)$ \cite{shaban2020nearfield} with ${\rm NF}$ being the noise figure in dBm and ${\rm BW}$ is the communication bandwidth in Hz

\begin{figure}

    \centering
\definecolor{mycolor1}{rgb}{0.00000,0.44700,0.74100}%
\definecolor{mycolor2}{rgb}{0.85000,0.32500,0.09800}%
\begin{tikzpicture}[scale=1\columnwidth/10cm]

\begin{axis}[%
width=8cm,
height=5cm,
at={(1.011in,0.642in)},
scale only axis,
xmin=25,
xmax=100,
xtick={ 25,  36,  49,  64,  81, 100},
xlabel style={font=\color{white!15!black}},
xlabel={Number of RIS Unit Elements $L$},
ymode=log,
ymin=0.0788746368099967,
ymax = 1,
ytick={0.1, 1},
yminorticks=true,
ylabel style={font=\color{white!15!black}},
ylabel={Positioning RMSE [m]},
axis background/.style={fill=white},
xmajorgrids,
ymajorgrids,
yminorgrids,
legend style={legend cell align=left, align=left, draw=white!15!black},
legend columns=2
]
\addplot [color=mycolor1, mark=o, mark options={solid, mycolor1},line width=1.0pt]
  table[row sep=crcr]{%
25	0.174293125100705\\
36	0.139580580195288\\
49	0.116135509147891\\
64	0.0928308402184045\\
81	0.0909059213941892\\
100	0.0837108406944358\\
};
\addlegendentry{DFT}

\addplot [color=mycolor2, mark=o, mark options={solid, mycolor2},line width=1.0pt, dashed]
  table[row sep=crcr]{%
25	0.20784339702774\\
36	0.180304438207342\\
49	0.13209657640226\\
64	0.11098710350464\\
81	0.0854201853545224\\
100	0.0788746368099967\\
};
\addlegendentry{Directive}

\addplot [color=mycolor1, mark=square, mark options={solid, mycolor1},line width=1.0pt]
  table[row sep=crcr]{%
25	0.521127751153846\\
36	0.259184295888563\\
49	0.183496552222617\\
64	0.208741627926607\\
81	0.106536635737883\\
100	0.100345882306094\\
};
\addlegendentry{DFT, $b=2$}

\addplot [color=mycolor2, mark=square, mark options={solid, mycolor2},line width=1.0pt, dashed]
  table[row sep=crcr]{%
25	0.501008372503908\\
36	0.334405521745377\\
49	0.258734285483262\\
64	0.221877934931354\\
81	0.109884415790844\\
100	0.0982707788761033\\
};
\addlegendentry{Directive, $b=2$}

\addplot [color=mycolor1, mark=triangle, mark options={solid, mycolor1},line width=1.0pt]
  table[row sep=crcr]{%
25	0.208339331383997\\
36	0.146989452960623\\
49	0.133324342824423\\
64	0.0982572404892799\\
81	0.127841515701769\\
100	0.0792885750062671\\
};
\addlegendentry{DFT, $b=3$}

\addplot [color=mycolor2, mark=triangle, mark options={solid, mycolor2},line width=1.0pt, dashed]
  table[row sep=crcr]{%
25	0.234273620166086\\
36	0.210618438108436\\
49	0.179012789530954\\
64	0.110504840590779\\
81	0.125521503579091\\
100	0.0906638227119801\\
};
\addlegendentry{Directive, $b=3$}

\end{axis}

\end{tikzpicture}%

    \caption{The RMSE of the positioning error in meters versus the number of RIS unit elements $L$, considering pilot symbol transmission with power $P=10$ dBm and the LS-based line intersection technique with $M=3$ RISs placed at $\mathbf{p}_1=[0\,5\,7]^{\rm{T}}$, $\mathbf{p}_2=[5\,0\,1]^{\rm{T}}$, and $\mathbf{p}_3=[10\,6\,8]^{\rm{T}}$. The user was located at the unknown position $\mathbf{p}=[4\,8\,2]^{\rm{T}}$. Different forms of the phase profiles for the RISs have been used.}\vspace{-4mm}
    		\label{fig:rmse}
\end{figure}
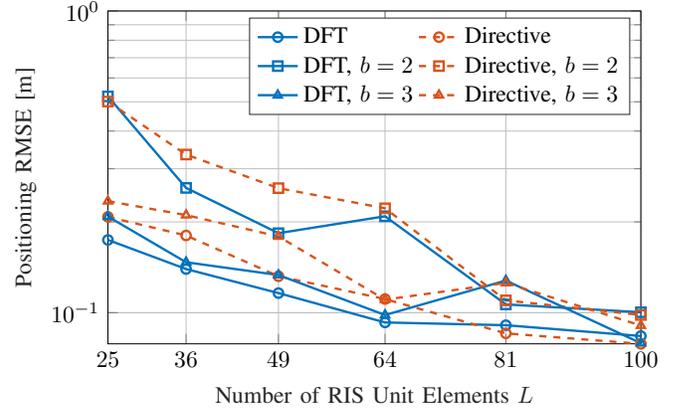
%\begin{figure}[!t]
%	\centering
%	\includegraphics[width=0.53\textwidth]{Figures/rmse_vs_num_el.pdf}
%	  \caption{The RMSE of the positioning error in meters versus the number of RIS elements $L$ considering source symbols transmitted with power $P=10$ dBm and the LS-based line intersection technique with $M=3$ RISs placed at $\mathbf{p}_1=[0\,5\,7]^{\rm{T}}$, $\mathbf{p}_2=[5\,0\,1]^{\rm{T}}$,  $\mathbf{p}_3=[10\,6\,8]^{\rm{T}}$ and the user positioned at $\mathbf{p}=[4\,8\,2]^{\rm{T}}$ . Different forms of the phase profiles for the RISs have been used.}\vspace{-0.2cm}
%	  		\label{fig:rmse}
%\end{figure}

\vspace{-2mm}
\subsection{Discussion}
\subsubsection{Impact of RISs' Placement on PEB and Coverage}
In Fig.~\ref{fig:cdf_plot}, the empirical CDF of the PEB for various source positions with $z_p=5$ ${\rm m}$ and different placements of $3$ RISs with $8\times8$ elements is illustrated. As shown, the placement of the RISs on the room's walls impacts the localization performance of the proposed AoA-based positioning approach. For best overall coverage, placing the RIS on different walls is better, while for best localization performance near a single wall, it is preferred to place the RIS on that same wall. The heights of the placements of the RISs on the wall(s) play only a minor role, though the case of same heights appears better for localization performance, while the case of different heights leads to better overall coverage.

%Among the considered cases, the placement in three different walls yields the larger probability for a PEB around $10^{-3}$ ${\rm m}$. 

\subsubsection{Impact of RISs' Size and Phase Profiles on LS Estimation}
The RMSE positioning performance, as defined in \eqref{eq:peb}, is plotted versus $L$ in Fig.~\ref{fig:rmse}, {using the proposed ML positioning technique} in \eqref{eq:ML} for the case of $3$ RISs and various forms for their phase profiles. Specifically, we have considered position estimation with: \textit{i}) $T=L$ phase profiles using the DFT beams {in \eqref{eq:dft} and \textit{ii}) quantized DFT beams with $b=2,3$ according to $\mathcal{F}$} ; \textit{iii}) $T=\lfloor L/3 \rfloor$ of the DFT phase profiles in \textit{i}) pointing around the actual direction of the source with $30^{\degree}$ uncertainty range; \textit{iv}) the $b$-bit quantized versions of the phase profiles in \textit{iii}). %
It is evident from the figure that, for all considered RIS phase profiles, the RMSE does not improve drastically with increasing $L$, and that the $T=L$ DFT profiles yield the best performance, with the directional $T\ll L$ DFT profiles requiring less overhead at a cost of a small performance loss. In addition, it is shown that quantization of both DFT and directive beams imposes performance loss, something that is less severe with higher $b$.

Hybrid schemes with progressive refinement of the user location could form an interesting compromise between performance (RMSE) and observation latency ($T$). 
%Interestingly, any prior knowledge about the source's signal AoA at each $m$-th RIS (which is implied by the use of the $\lfloor L/3 \rfloor$ directional phase profiles of the \textit{i}) and \textit{ii}) cases), improves the RMSE, approaching the performance with the $L$ DFT profiles. In fact, as $b$ increases, the performance with quantized directional profiles improves.
\begin{figure}
    \centering
\definecolor{mycolor1}{rgb}{0.63500,0.07800,0.18400}%
\definecolor{mycolor2}{rgb}{0.00000,0.44700,0.74100}%
\begin{tikzpicture}[scale=1\columnwidth/10cm]

\begin{axis}[%
width=8cm,
height=5cm,
at={(1.011in,0.642in)},
scale only axis,
xmin=-15,
xmax=15,
xlabel style={font=\color{white!15!black}},
xlabel={Transmit Power [dBm]},
ymode=log,
ymin=0.02,
ymax=6.652,
yminorticks=true,
ylabel style={font=\color{white!15!black}},
ylabel={Positioning RMSE [m]},
axis background/.style={fill=white},
xmajorgrids,
ymajorgrids,
yminorgrids,
legend style={legend cell align=left, align=left, draw=white!15!black},
legend columns=2,
legend pos=north east
]
\addplot [color=mycolor1, line width=1.0pt]
  table[row sep=crcr]{%
-15	0.458550273073984\\
-10	0.257861768226820\\
-5	0.145006328460596\\
0	0.0815430509075197\\
5	0.0458550273073983\\
10	0.0257861768226821\\
15	0.0145006328460598\\
};
\addlegendentry{\small{PEB, 4 RISs}}

\addplot [color=mycolor2, line width=1.0pt, dashed]
  table[row sep=crcr]{%
-15	0.706582337931950\\
-10	0.397340448268749\\
-5	0.223440954231175\\
0	0.125650082304156\\
5	0.0706582337931950\\
10	0.0397340448268748\\
15	0.0223440954231175\\
};
\addlegendentry{\small{PEB, 3 RISs}}

\addplot [color=mycolor1, mark=o, line width=1.0pt, mark options={solid, mycolor1}]
  table[row sep=crcr]{%
-15	5.35476977775404\\
-10	2.49177220647986\\
-5	0.161278996251825\\
0	0.0899918137340033\\
5	0.0691371995170080\\
10	0.0546943676512246\\
15	0.0492382968278465\\
};
\addlegendentry{\small{ML, 4 RISs, FF}}

\addplot [color=mycolor2, mark=o, line width=1.0pt, mark options={solid, mycolor2}, dashed]
  table[row sep=crcr]{%
-15	6.45396874811166\\
-10	3.34883443895480\\
-5	0.232920366618123\\
0	0.120927051591549\\
5	0.0687626214701920\\
10	0.0406078702284715\\
15	0.0246348498552220\\
};
\addlegendentry{\small{ML, 3 RISs, FF}}

\addplot [color=mycolor1, mark=square, line width=1.0pt, mark options={solid, mycolor1}]
  table[row sep=crcr]{%
-15	3.57998629901968\\
-10	1.68862324088331\\
-5	0.319052260058739\\
0	0.131978881321292\\
5	0.0932485709615470\\
10	0.0725237711316077\\
15	0.0708697357435194\\
};
\addlegendentry{\small{ML, 4 RISs, NF}}

\addplot [color=mycolor2, mark=square, line width=1.0pt, mark options={solid, mycolor2}, dashed]
  table[row sep=crcr]{%
-15	6.65200688512510\\
-10	3.43469899092000\\
-5	0.236562388179611\\
0	0.147271673914487\\
5	0.118194626290541\\
10	0.0918055378796242\\
15	0.0875944633506508\\
};
\addlegendentry{\small{ML, 3 RISs, NF}}

\end{axis}

\end{tikzpicture}%
    \caption{The RMSE of the positioning error in meters versus the source transmit power $P$ in dBm considering $M=3$ and $4$ RISs, each with $L=8\times8$ unit elements, placed at $\mathbf{p}_1=[0\,5\,7]^{\rm{T}}$, $\mathbf{p}_2=[5\,0\,1]^{\rm{T}}$,  $\mathbf{p}_3=[10\,6\,8]^{\rm{T}}$, and
	  $\mathbf{p}_4=[4\,10\,6]^{\rm{T}}$. The user was positioned at {$\mathbf{p}=[4\,8\,2]^{\rm{T}}$ and $64$ quantized DFT phase profiles were used for each RIS. The proposed ML-based position technique was evaluated along with the PEB, considering both the Far-Field (FF) and Near-Field (NF) channel models.}} \vspace{-4mm}
    		\label{fig:tr_power}
\end{figure}
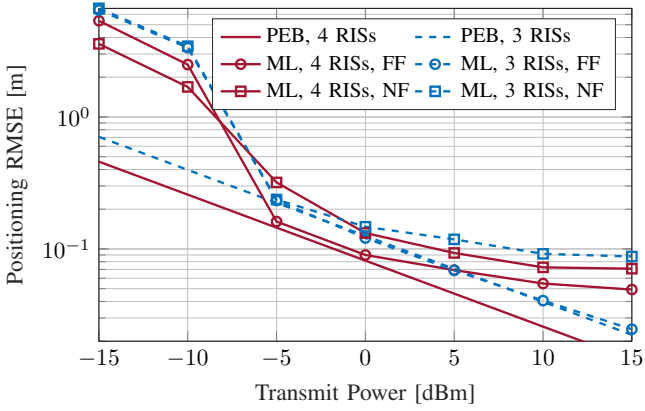

%\begin{figure}[!t]
%	\centering
%	\includegraphics[width=0.53\textwidth]{Figures/peb_ml_ls_3_vs_4_ris.pdf}
%	  \caption{The RMSE of the positioning error in meters versus the source transmit power $P$ in dBm considering $M=3$ and $4$ RISs each with $L=8\times8$ elements placed at $\mathbf{p}_1=[0\,5\,7]^{\rm{T}}$, $\mathbf{p}_2=[5\,0\,1]^{\rm{T}}$,  $\mathbf{p}_3=[10\,6\,8]^{\rm{T}}$,
%	  $\mathbf{p}_4=[4\,10\,6]^{\rm{T}}$
%	  and the user positioned at $\mathbf{p}=[3\,7\,5]^{\rm{T}}$ and $21$ random phase profiles. Both the proposed ML-based and LS-based position techniques were evaluated along with the PEB, considering both the far-field and near-field channel models.}\vspace{-0.2cm}
%	  		\label{fig:tr_power}
%\end{figure}

\subsubsection{Impact of Number of RISs and Channel Model on Estimation Performance}
The RMSE of the positioning error as a function of $P$ in dBm, considering $M\in \{3,4\}$ RISs,  each with $L=8\times8$ elements and {$T=64$ quantized DFT phase profiles, is illustrated in Fig.~\ref{fig:tr_power}. We have considered the ML positioning technique with the far-field and near-field channel models. In addition, PEB curves are also included under the far-field model. It can be observed that, as expected, the PEB improves with increasing $P$ and $M$. Moreover, the ML-based technique performs sufficiently close to the PEB from low $P$ values. It is also shown for the latter technique that, for $P$ larger than $0$ dBm and the near-field channel model, the RMSE saturates around $0.1$ ${\rm m}$ for both $3$ and $4$ RISs and, for the $4$-RIS case with the far-field channel model, the RMSE saturates around $0.07$ ${\rm m}$. This shows that the ML technique exhibits an error floor due to the model mismatch in the case of the near-field, and due to multipath for the far-field case, but is still able to provide good positioning quality.}

\section{Conclusion}%\vspace{-0.15cm}
In this paper, we presented an ML-based localization method with multiple single-RX-RF RISs, which relies on a beamspace OMP technique for AoA estimation and LS-based line intersection. The presented numerical results showcased the accuracy of the proposed method under reasonable operating conditions for strong LOS multipath channels and single-source positioning, verifying our theoretical estimation analysis. We intend to extend the presented localization framework to hybrid RISs \cite{HRIS} and multiple mobile active users. 

\appendix%\vspace{-0.15cm}
{Using the notations $\nu\triangleq\phi_{m,1}$ and $\sigma\triangleq\theta_{m,1}$, it follows from the definition of the function ${\mu}_{m,t}$ in \eqref{eq:FIM_Initial} that:}
\begin{equation}
\frac{\partial {\mu}_{m,t}}{\partial\boldsymbol{\eta}_m}=\begin{bmatrix}
\mathbf{u}_{m,t}^{\rm H}\boldsymbol{\alpha}\left(\nu,\sigma\right)\\
\jmath\sqrt{P_{L_{m,1}}}\mathbf{u}_{m,t}^{\rm H}\boldsymbol{\alpha}\left(\nu,\sigma\right)\\
\sqrt{P_{L_{m,1}}}\mathbf{u}_{m,t}^{\rm H}\frac{\partial\boldsymbol{\alpha}\left(\nu,\sigma\right)}{\partial\nu}\\
\sqrt{P_{L_{m,1}}}\mathbf{u}_{m,t}^{\rm H}\frac{\partial\boldsymbol{\alpha}\left(\nu,\sigma\right)}{\partial\sigma}
\end{bmatrix}\exp\left(\jmath\varphi_{m}\right)s,
\end{equation}
where the included partial derivatives are derived as: 
\begin{equation}\frac{\partial\boldsymbol{\alpha}\left(\nu,\sigma\right)}{\partial\tau}=\exp\left(-\jmath\mathbf{q}^{\rm T}_{m}\mathbf{k}(\nu,\sigma)\right)\odot\jmath\mathbf{q}^{\rm T}_{m}\frac{\partial\mathbf{k}(\nu,\sigma)}{\partial\tau}
\end{equation}
with $\tau$ being either $\nu$ or $\sigma$ and $\mathbf{q}_{m}\triangleq[\mathbf{q}_{m,1}\,\mathbf{q}_{m,2}\,\cdots\,\mathbf{q}_{m,L}]$. The derivatives with respect to $\nu$ and $\sigma$ are computed as:
\begin{equation}
\frac{\partial\mathbf{k}(\nu,\sigma)}{\partial\nu}=-\frac{2\pi}{\lambda}\begin{bmatrix}
-\sin{\sigma}\cos{\nu}\\
\sin{\sigma}\sin{\nu}\\
0
\end{bmatrix},
\end{equation}
\begin{equation}
\frac{\partial\mathbf{k}(\nu,\sigma)}{\partial\sigma}=-\frac{2\pi}{\lambda}\begin{bmatrix}
\cos{\sigma}\cos{\nu}\\
\cos{\sigma}\sin{\nu}\\
-\sin{\sigma}
\end{bmatrix}.
\end{equation}

The Jacobian $\mathbf{t}_m$ required in \eqref{eq:fim_position} is calculated, as follows. We express the elevation angle as a function of the coordinates:  
\begin{equation}
\sigma=\arccos\left(\dfrac{z_p-z_m}{\left\|\mathbf{p}-\mathbf{p}_m\right\|}\right),
\end{equation}
and then calculate the required partial derivative as:
\begin{equation}
\frac{\partial\sigma}{\partial\mathbf{p}} \triangleq \begin{bmatrix}
\frac{\partial\sigma}{\partial x_p}\\
\frac{\partial\sigma}{\partial y_p}\\
\frac{\partial\sigma}{\partial z_p}
\end{bmatrix}= \begin{bmatrix}
\left(z_p-z_m\right)\left(x_p-x_m\right)(\kappa\chi)^{-1}\\
\left(z_p-z_m\right)\left(y_p-y_m\right)(\kappa\chi)^{-1}\\
\left(\frac{\left(z_p-z_m\right)^2}{\chi}-\frac{1}{\left\|\mathbf{p}-\mathbf{p}_m\right\|_2}\right)\kappa^{-1}
\end{bmatrix},
\end{equation}
where $\kappa\triangleq\sqrt{1-\frac{\left(z_p-z_m\right)^2}{\left\|\mathbf{p}-\mathbf{p}_m\right\|^2}}$ and $\chi\triangleq\left\|\mathbf{p}-\mathbf{p}_m\right\|^3$. The partial derivative of $\nu$ is obtained in a similar manner.

\bibliographystyle{IEEEtran}
\bibliography{references}

\end{document}